\documentclass{ws-procs975x65}

\begin{document}

\title{Gravitons production in a new type of GCG
}

\author{Mariam  Bouhmadi-L\'opez$^1$, Pedro Fraz\~ao$^2$ and Alfredo B. Henriques$^3$}

\address{Centro Multidisciplinar de Astrof\'{\i}sica - CENTRA,
 Departamento de F\'{\i}sica,\\ Instituto Superior T\'ecnico,
Av. Rovisco Pais 1, 1096 Lisboa, Portugal\\E-mails: $^1$ mariam.bouhmadi@ist.utl.pt, $^2$ pedro.frazao@ist.utl.pt, $^3$ alfredo.henriques@ist.utl.pt}

\begin{abstract}
A new scenario for the early era of the Universe is proposed. It corresponds  to a smooth transition between a de Sitter-like phase and a radiation dominated era. We calculate the production of gravitons in this model.
\end{abstract}


\bodymatter

\section{Motivation}

Inflation is the most consistent model to explain the anisotropies in the cosmic microwave background radiation (CMBR), and also the origin of the large scale structure (LSS) of the Universe.
As the Universe evolves, there is, in addition, a production of  a stochastic  background of primordial gravitational waves (GW), originated from the vacuum fluctuations.

In what follows, we set a phenomenological model for unifying the transition from the inflationary era to the subsequent radiation dominated epoch of the Universe. In particular, we consider a way to extend the framework of the generalised Chaplygin gas (GCG)\cite{chaplygin} to the primordial evolution of the universe. 

This transition is not well understood and it is important to investigate the possible signatures in the power spectrum of the stochastic background of GW associated with it, giving a new insight on the inflationary model behind the accelerated expansion in the early universe.

\section{A GCG inspired scenario for the first stages of the Universe}

This unification is achieved by considering a perfect fluid whose energy density, $\rho$, has a dependence on the scale factor, $a$,  of the form \cite{work}
\begin{equation}
\rho=\left(A+\frac{B}{a^{4(1+\alpha)}}\right)^{\frac{1}{1+\alpha}}\;,
\end{equation} 
where $A,B$ are constants determined by the matter content of the universe, and $\alpha$ is a free parameter. The conditions, $1+\alpha<0$, and $A,B>0$, lead to the expected interpolation.
The conservation of $\rho$  implies the following
 equation of state  
\begin{equation}
 P=\frac13\rho -\frac43\frac{A}{\rho^{\alpha}}\;.
\end{equation}

\section{The Inflationary Scalar Field}

The inflationary dynamics of this model can be described by a scalar field, $\phi$, rolling down its self-interacting potential, $V(\phi)$. This potential can be obtained by identifying $\rho$ and $P$ of the fluid with those of the scalar field.
On the other hand, the number of e-folds of expansion since a given mode $k$ exits the horizon ($k=aH$) during the inflationary era, at $\phi=\phi_c$, until the end of inflation can be approximated by  $N_c\thickapprox \kappa^2\int_{\phi_{\star}}^{\phi_c}\frac{V}{V_\phi}\;d\,\phi$.

\section{Observational Constraints}

The model can be constrained using CMBR/LSS recent measurements. More precisely,  with the values\cite{Lidsey97} $P_s(k_c)=(2.45\pm 0.23)\times 10^{-9}$ and  
$n_s=1.0\pm 0.1$ for the power spectrum and spectral index of the scalar perturbations, respectively,
 evaluated at the pivot wave number $k_c=0.05\;\mathrm{Mpc^{-1}}$.

Therefore, the energy scale for inflation can be determined through the  amplitude of the power spectrum for density perturbations, 
$P_{s}(k) \approx \frac{\kappa^6}{12\pi^2}\frac{V^3}{V_\phi^{2}} \Big|_{k=aH}$,
where $V_\phi\equiv {dV\over d\phi}$. 
On the other hand, we can analyse the variation of the scalar to tensor ratio, $r \equiv \frac{P_t(k)}{P_s(k)} = 16\,\epsilon$, where $P_{t}(k)$ is the tensorial power spectrum, 
in terms of the spectral index of the scalar perturbations\cite{work}  $n_s=1-6\,\epsilon+2\,\eta$.   

There are two approaches to evaluate $r$ and $n_s$: one using the number of e-folds, $N_c$, and the other one by means of the comoving wavenumber, $k=aH$. The results are presented on the $n_s-r$ parameter space in Fig.~1-a, where we see that the parameter space is slightly more red than preferred by the observation (blue area). However, Fig.~1-a also shows that $r<1$ which is in agreement with the observation\cite{Lidsey97}. The most consistent value of $\alpha$ with the observation is $-1.024$,  corresponding to an energy scale for inflation of  $\sim 10^{16}$ GeV. 

\section{Gravitational Wave Spectrum}

The GW spectrum can be calculated using the method of the continuous Bogoliubov coefficients, $\alpha$ and $\beta$\cite{abh1}. 
The dimensionless relative logarithmic energy of the spectrum of the GWs, $\Omega_{\rm GW}$, at the present time $\eta_0$, is related to\cite{abh1} $\beta$: 
\begin{equation}
\Omega_{\mbox{\scriptsize GW}} = \frac{\hbar \kappa^2}{3\pi^2 c^5 H^2(\eta_0)} \omega^4 \beta^2\;,
\end{equation} 
where $\beta^2$ is proportional to the number of gravitons created and $\omega$ is the respective angular frequency.
The same coefficient can be written in terms of two continuous functions of time, $X$ and $Y$, which obey the differential equation\cite{abh1}
\begin{equation} X^{\prime\prime } +\frac{i}{k} \left( k^{2}-\frac{a^{\prime \prime }}{a} \right) X=0\;,\label{GWeq}\end{equation}
 with $X^\prime = -i k Y$.  As initial conditions for the numerical integration of Eq.~(\ref{GWeq}); i.e. the initial values for $X(\eta)$ and $Y(\eta)$, we use the exact analytical solution for a dS Universe\cite{abh1} because our model is quasi-Sitter in the past.
For the late-time evolution we use the LCDM model defined through the  recent measurements of WMAP5.

In Fig.~1-b we present the GW spectra  resulting from the numerical integrations using the number of e-folds to constrain the inflationary scale. As can be noticed, the spectrum changes with $\alpha$, with a simultaneous vertical displacement and a variation in the high frequency region.

The decrease of the plateau height, which can be observed by the difference between the solid lines ($N_c=47$) and the dashed lines ($N_c=62$) for a constant $\alpha$ (a given colour), is easily interpreted as a decrease in the inflationary scale.
On the other hand, we immediately conclude that the increase in the high frequency region, when $\alpha$ moves away from $-1$,  is a function of the parameter $\alpha$ only.
\begin{figure}[h]
  \centering
  \includegraphics[width=6cm]{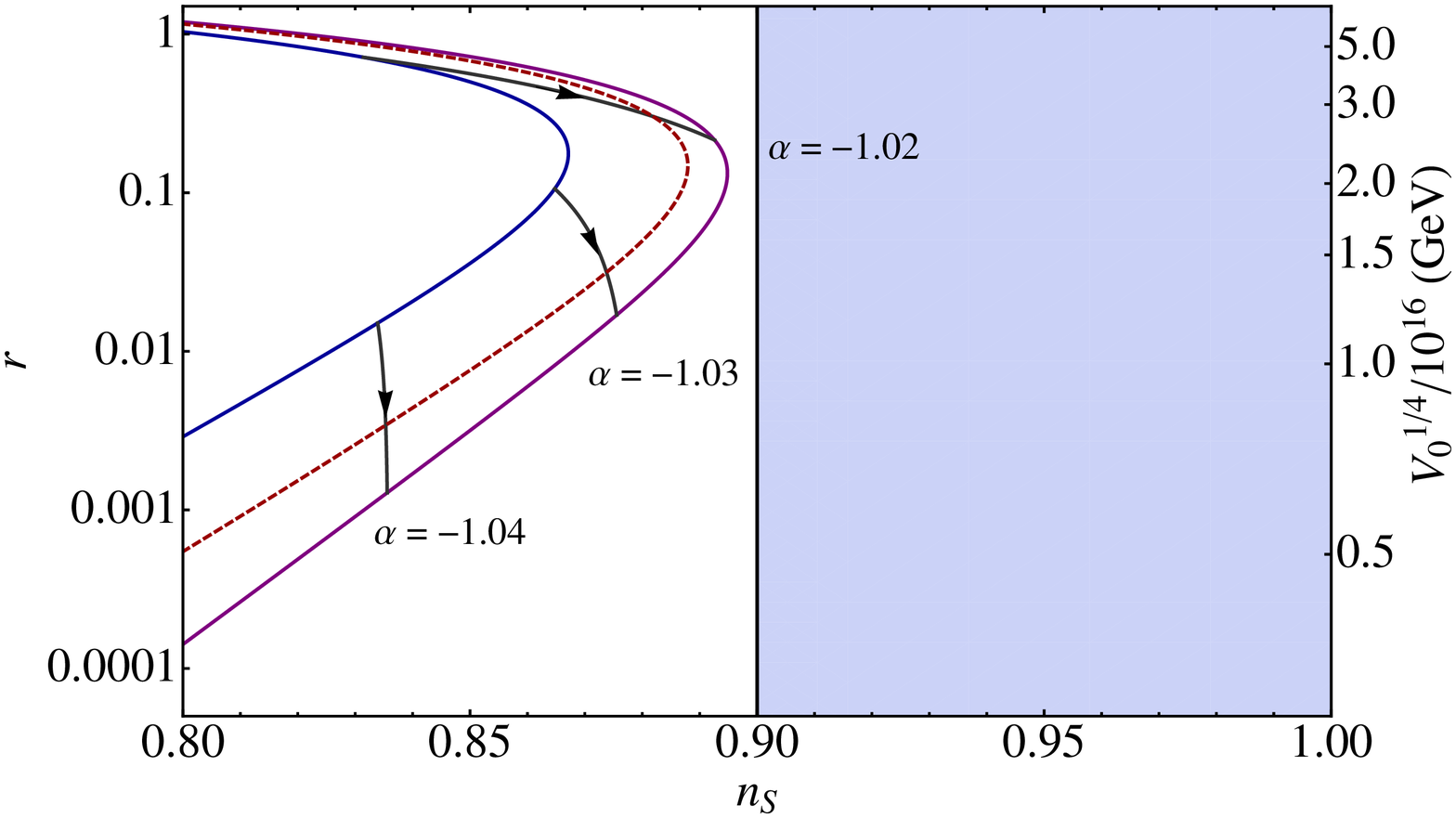}\hspace{0.4cm}\includegraphics[width=5.4cm]{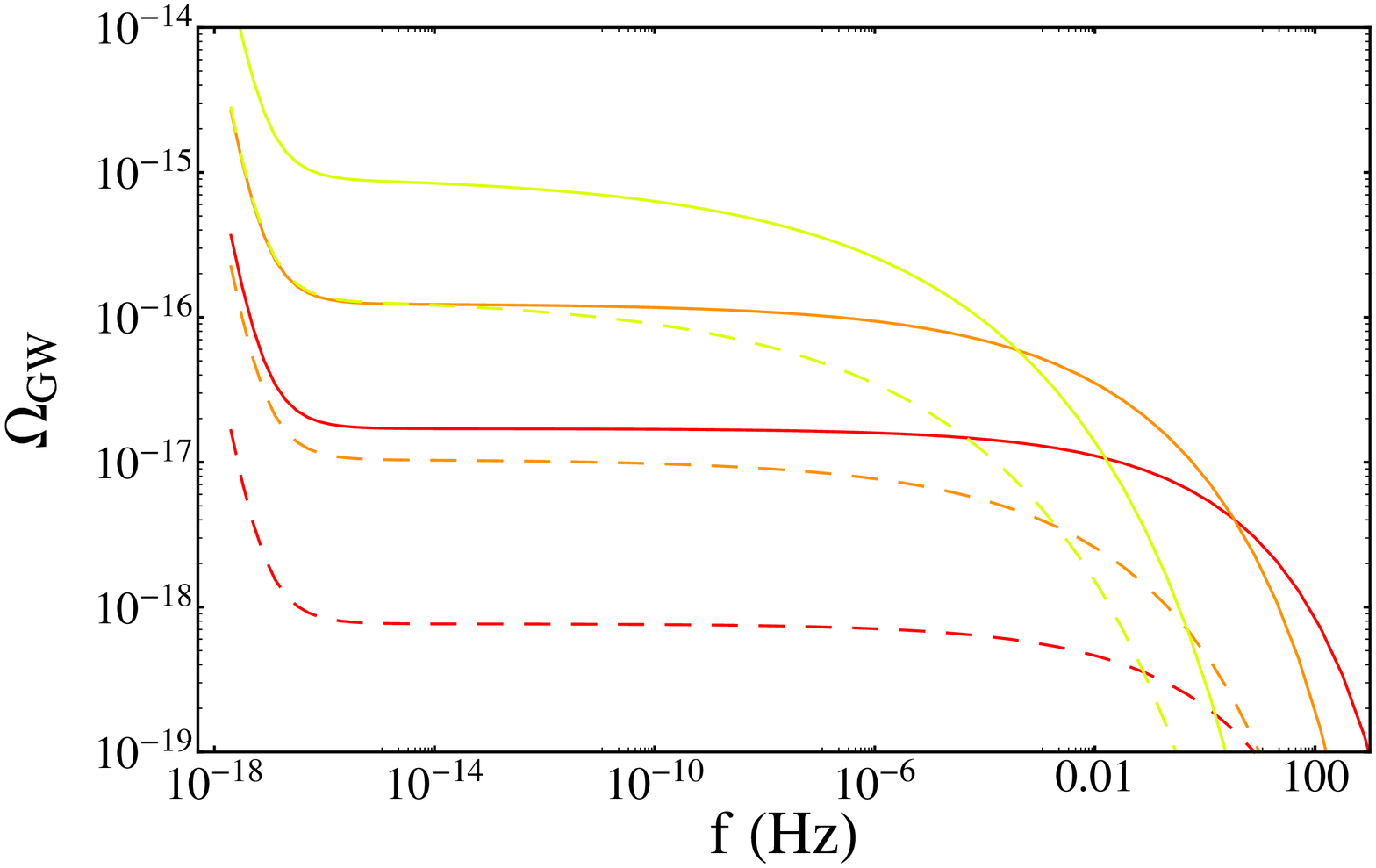}
  \caption{Fig. 1-a (left) presents the resulting $n_s-r$ parameter space, using the number of e-folds of $N_{\rm c}=47$ in blue, and $N_{\rm c}=62$ in purple, or from the comoving wavenumber, $k$, in red. The black lines are for a constant $\alpha$-value, parameterised by an increase in the number of e-folds. Fig. 1-b (right) shows the obtained gravitational wave spectra.}
  \label{results}
\end{figure}

\vspace{-0.6cm}

\section{Conclusions}

The GW spectra reveal a consistent picture
corresponding to the described smooth transition, 
and represents a significant signature of this modified generalised Chaplygin gas.
The strong variation of the high frequency range, is directly related with the decrease in the maximum of the potential $a^{\prime\prime}/a$, when $\alpha$ approaches $-1$. 
This issue shows the strong limits to the maximum frequency accepted in our model and which will be within the reach 
of future gravitational-waves detectors like BBO and DECIGO\cite{Lidsey97}, for the 
Khz range of frequencies. In fact, for the most consistent values of the $\alpha$-parameter, the spectrum
shows a frequency as low as in the Hz region.

\vspace{-0.3cm}\section*{Acknowledgements}
\vspace{-0.2cm}M.B.L. is supported by FCT (Portugal) through the fellowship SFRH/BPD/26542/2006.

\end{document}